# ARTICLE

# Novel one-step electrophoretic deposition of membrane-electrode assembly for flexible-batteries application

Elazar Cohen[a], Moran Lifshitz[a], Alexander Gladkikh[b], Yossi Kamir[a], Ido Ben-Barak[a] and Diana Golodnitsky[a,b]*

**Wearable electronic devices and gadgets raise the need for flexible, thin and lightweight batteries. In this article we present for the first time, a unique, single-step method for the preparation of a membrane-electrode assembly for flexible-batteries application. Concurrent electrophoretic deposition (EPD) of positive and negative battery electrodes (LFP and LTO) on opposite sides of a commercial nanoporous membrane (Celgard 2325) results in the formation of a three-layer-battery structure. The cell comprising this electrophoretically deposited structure ran for more than 150 cycles with 125-140mAh/g capacity, which approaches the theoretical value of lithium iron phosphate. The electrodes can be deposited either cathodically or anodically by replacing the interchangeable charging agents, like polyethyleneimine and polyacrylic acid. These polyelectrolytes, when adsorbed on the particles of the active material, serve also as the binders. The simultaneous EPD, which we developed, can be used for the simple and low-cost manufacturing of a variety of cathode and anode materials on nanoporous polymer- and ceramic ion-conducting membranes for energy storage devices.**

## Introduction

The forthcoming era of smart and omnipresent energy requires flexible power sources, which feature visual adaptability and simplistic integration with electronic devices of various sizes and shapes [1]. The growing investments in the development of wearable technologies and devices are significantly boosting the revenue generation opportunities for flexible battery manufacturers. The focus on shifting towards miniaturized products coupled with the booming demand for consumer electronics are some of the key driving factors behind the flexible-battery market. In the development of innovative power sources, deliverance from design limitation, as well as the synthesis of high-performance electrochemical materials with well-tuned features, is considered to be the most important technical criterion.

Research on flexible lithium-ion batteries has mainly been focused on the development of nano-size electroactive materials, shape-conformable electrolytes, and soft current

a- School of Chemistry, b- Wolfson Applied Materials Research Center, Tel Aviv University, Tel Aviv, 6997801, Israel
*corresponding author: Prof. Diana Golodnitsky, FRSC, email: golod@tauex.tau.ac.il

collectors. Free-standing carbon-based mats based on carbon nanotubes (CNTs), graphene, and their composite matrices have attracted increasing attention as promising functional electrode materials for flexible batteries [1]. Cui et al. [2] presented a new thin, flexible lithium-ion battery using plain-paper separator and free-standing thin films of carbon nanotubes as current collectors. A lamination process was used for the integration of the current collectors, $Li_4Ti_5O_{12}$ and $LiCoO_2$ battery materials onto a single sheet of paper. The 300μm-thick lithium-ion paper battery had robust mechanical flexibility and delivered 108mAh.g$^{-1}$ reversible capacity. An additional lamination process was required to apply a thin layer of PVdF on the paper in order to prevent the internal shorts of the device, caused by leakage of the electrode material through a large hole in the paper. In [3] the robust 30μm-thick network architecture, comprising interpenetrating nanocomposites of ultra-long carbon nanotubes and vanadium pentoxide nanowires, which enables 169mAh.g$^{-1}$ capacity at high charge/discharge rates was presented. Lithium metal was used as an anode in this battery. CNT tissue, as a promising candidate for the anode current collector in lithium-ion batteries, has been proposed by Ein-Eli et al. [4] Graphene-paper-based electrodes have been tested in flexible battery designs, as well [5, 6]. The integration of graphene into the $V_2O_5$ electrode enabled the fabrication of thin, lightweight, and flexible batteries with much higher capacity than obtained in batteries assembled on conventional current collectors [5]. A comprehensive review of electrochemical energy-storage devices for wearable technology has been recently published by Liu et al. [7]. Wearable planar electrochemical energy-storage devices have been discussed with the focus on recent advances in the materials, cell designs, fabrication methods and electrochemical performance during mechanical deformation [7-10]. In fact, the procedure of processing of the powdered material to obtain thin or thick mechanically strong films on the preferred current collector appears to be more complicated than the choice of the positive or the negative electrode material to fabricate a flexible lithium-ion battery. In the current studies, doctor blade technique, chemical vapor deposition, vacuum-assisted filtration, sputtering and nanofabrication are usually exploited for the preparation of flexible electrodes [10]. [10]. These processes are reasonable in lab-scale studies, but they are insufficient for scaling up to mass production. Physical vapor deposition (DC or RF magnetron sputtering) and pulsed laser-deposition techniques typically used for depositing electrode materials for flexible batteries are quite expensive. In addition, their top-down deposition nature does not meet the need to conformally deposit highly adhesive electrode materials on current collectors [11]. CVD techniques are expensive and heating of the substrate during deposition or annealing of the obtained coating is usually required to ensure crystallinity for optimum battery performance [12, 13].

For quite a while, solution-based processing of electrode materials, which include ink-jet printing, serigraphy, sol–gel, chemical-bath, electro- and electrophoretic deposition techniques, has been shown to be accessible and cost-effective for the fabrication of thin-film high-performance electrodes [11, 14-22]. Electrophoretic deposition (EPD) is one of solution-based techniques, which currently gained much attention for formulating battery electrodes. This is due to the high versatility of materials types, particle size and crystallinity of the starting powders preserved in the deposited films [11, 22-24]. EPD provides good conformal deposits on complicated geometrical surfaces and is cost-effective method. The thickness of these conformal films and mass loading of electrophoretically deposited materials can be easily controlled by varying the colloidal-electrolyte composition, the applied voltage or current, and deposition time. We have recently presented [22] electrophoretically deposited thin-film composite LiFePO4 cathodes and studied the effect of polymers, surface-active additives and deposition parameters on the adhesion, compactness and electrochemical performance of LiFePO4 films in planar and three-dimensional microbatteries. Several groups reported on preparation of LMP (lithium manganese phosphate), LCO (lithium cobalt oxide), LMO (lithium manganese oxide) cathodes and graphite, LTO (lithium titanate) and silicon anodes for lithium-ion batteries by electrophoretic deposition [25-28]. Aiming at the improvement of electrochemical properties of electrodes, the modification of inter-particle connectivity and packing density by compressing the deposited films was proposed [29, 30]. Controlling EPD electrolyte recipes to form $Co_3O_4$/graphene sandwich-like layered structure was reported in [26, 31], and improvement of cycling performance by annealing of EPD electrode – in [32]. Our group pioneered EPD formation of tri-layered battery structures, comprising LiFePO4 cathode, LiAlO2-PEO (lithium aluminate-polyethylene oxide) or $Li_{(1-x)}Al_yGe_{(2-y)}(PO_4)_3$-PEI (LAGP, lithium aluminum germanium phosphate – polyethylene imine) membrane, and LiTiO$_2$-based anode, on conductive graphene substrates [33].

To the best of our knowledge, very few articles address the electrophoretic deposition of ceramic materials on polymer membranes. The first report is that of Sarkar and Nicholson in 1996 [34], who used this approach to deposit cathodically yttria-stabilized zirconia (YSZ)/Al$_2$O$_3$ micro-laminates in ethanol suspension. They used a dialysis membrane in order to separate the anode and the cathode compartments in a suspension and reported on the formation of layer-by-layer coating on the membrane. Ordung et al [35] presented deposition of homogeneous silicon-powder-based film on fibre fabrics, with a high packing density of more than 60vol% and good mechanical properties.

The feasibility of preparation by EPD of a membrane electrode assembly (MEA) for a polymer-electrolyte fuel cell was demonstrated in [36]. The deposition of carbon with a platinum catalyst on a Nafion® polymer substrate was confirmed by SEM observation and XPS analysis.

The possibility of simultaneous electrophoretic deposition of oppositely charged MgO and silicon particles was tested in [37]. MgO particles gain positive surface charge in acetone suspension, and are deposited on metal cathodes, while silicon particles have negative $\zeta$ potential and their deposition is anodic. Upon mixing of the two suspensions, the coulombic forces bring the particles together to form clusters. It was found that, depending on the relative contents of the powders, the deposition of clusters, comprising both, positively and negatively charged particles, occurs either on the cathode or the anode. The particles could not be separated and could not be deposited as a single-compound film on oppositely charged electrodes even under the strong applied electric field of 200V/cm.

Preparation of positive and negative battery electrodes by a one-step procedure is a very promising, but challenging approach. We present in this article for the first time, a unique fabrication method of three-layer battery assembly by one-step concurrent electrophoretic deposition of two battery electrodes directly on each side of a nanoporous membrane for flexible-battery application.

## Results and discussion

As declared in the Introduction, the goal of this research was the development of a membrane-electrode battery assembly (MEBA) by the simultaneous electrophoretic deposition of negative and positive battery electrodes on the opposite sides of a flexible nanoporous membrane.

Since most of the of the publications report the EPD of different materials on metal electrodes, like nickel and aluminum, or graphite, it was of particular importance to initially test the feasibility of electrophoretic deposition of the LFP and the LTO on a polymer membrane. Celgard 2325 membrane is a tri-layer polypropylene-polyethylene-polypropylene 25μm-thick film with a porosity of 39% and pore size not exceeding 28nm. This membrane, in addition, is known to have good wettability by acetone and by typical nonaqueous electrolytes used in lithium-ion batteries. In most of our experiments, the LFP was deposited cathodically and the LTO anodically on the membrane, which was placed in front of the electrode, both at 100V and from 30 to 300 seconds.

It is well established that the surface charge of particles dispersed in suspension is brought about by the adsorption of ionic species and/or polyelectrolytes Zhitomirsky et al showed [38] that PAA (polyacrylic acid) can be used for effective electrophoretic deposition of MnO2, NiO, TiO2, SiO2 and MWCNT materials. The authors state that this is the deprotonated COOH groups of the macromolecules in the solutions at pH>4, which induces the negative charge of PAA. The positive charge of PEI, the second polyelectrolyte used by us in the acetone-based suspension, is induced by the acetylacetone additive via a keto-enol reaction. We have recently found that linear PEI, when adsorbed on the surface, creates strong electrostatic repulsion between the particles, which in some cases even eliminates the formation of deposits. Branched PEI facilitates stabilization of the suspension, presumably via the electrosteric mechanism and promotes the formation of cathodic coatings.

The adhesion of electrophoretically deposited electrodes to Celgard 2325 was evaluated by an adhesive tape test. The tape was stuck on the film and peeled by pulling one of its ends. It was found that after the tape test the remaining area of the membrane coated by cathodic LFP deposit, in which PEI was used as a single charging agent/binder, was only half of that of the EPD film prepared from the suspension containing two polyelectrolytes - PEI and PAA.

It is important to emphasize that simply by replacing the charging agents in the appropriate suspensions, the LFP can be electrophoretically deposited on the anode and the LTO on the cathode. Moreover, it appears that LFP particles dispersed in acetone solvent with acetylacetone alone undergoes anodic deposition, while the ξ-potential induced by this additive to the LTO particles is insufficient to enable their EPD. Increased concentration of PAA (3% w/w) facilitates the migration of both types of particles towards the anode, while the combination of PEI (1% w/w) and PAA (2% w/w) – towards the cathode.

The mass of the deposited materials was found to increase with deposition time, leading to the formation of films of different thicknesses. Figure 1 shows that the electrophoretic deposition of both LFP and LTO particles occurs more quickly on a nickel substrate than on Celgard membrane. While the cathodic yield on nickel is higher than the anodic for both materials, the inverse dependence is seen for EPD on the membrane, namely, the cathodic process is much slower than the anodic. Moreover, we have found that at constant deposition voltage, insertion of a nanoporous polymer membrane perpendicularly to the electric field between the electrodes causes the current drop. This indicates that while there is a continuous path of protons through the micropores, their partial blockage by the deposit, followed by the increased resistance of the polymer partition, impedes the deposition rate. At present, there are insufficient data to decide whether, in the low-ionic-strength suspension, the major current is carried by the free ions, or if the charged particles are the primary charge carriers. A complex interplay between these two phenomena is most likely to occur.

Partial blockage, in addition, may result in the incomplete utilization of the active electrode material. The choice of appropriate particle size and polymer, membrane porosity and tortuosity, as well as, optimization of the EPD process can resolve this issue.

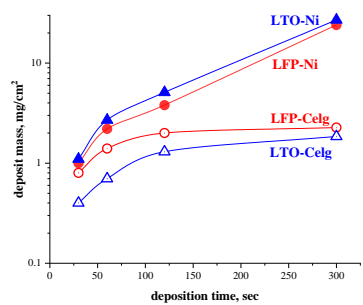

**Fig.1**. Deposition rates of cathodic electrophoretic deposition of LTO and anodic electrophoretic deposition of LFP on Ni and Celgard membrane

The surface of aluminum-foil electrodes after finishing the deposition on in-front-placed membrane was found to be completely clean with no traces of LFP- or LTO-based coatings. We ascribe this to the coagulation of the micelles, composed of LFP/LTO and polymers in the very close vicinity of the membrane and their precipitation on Celgard, the pores of which are smaller than the particle sizes of LFP and LTO.

As expected, when two and three Celgard membranes are placed in front of the negative aluminum electrode, the mass of the LFP deposit on the membrane facing the suspension, significantly decreases. The decrease is more pronounced than that estimated from the simple increase of the thickness of the dielectric barrier. We attribute this not only to the diminution of the applied electric field strength, but mainly to the imperfect contact between the membranes, which results in the nonlinearity of the external field.

While the mechanism of EPD has been the subject of much study, a full understanding of the process is still lacking [39]. The prevailing view today is that EPD involves electrophoresis and film formation. Zhitomirsky [40] proposed three major steps in the electrophoretic film formation, namely, charge-neutralization followed by electrocoagulation, $\zeta$-potential lowering, causing electrochemical coagulation, and particles accumulation. We agree with Grillion et al [41], that charge neutralization can occur when the particles undergo a redox reaction in contact with the electrode. It our experiments, this does not apply, since the particles are charged by polyelectrolyte macromolecules, which do not penetrate the membrane. Lowering of the $\zeta$ potential involves reduction of the repulsive forces between particles as a result of increase of ionic strength around the particles, change of the local pH and/or thinning of the part of the lyosphere adjacent to the electrode [39, 42]. For the deposition on the porous membrane the two mentioned reasons for $\zeta$-potential drop are improbable, however charge redistribution in the diffuse part of the polarizable particle's EDL, induced by the external electric field, cannot be excluded as an intermediate step. Hamaker and Verwey [43] suggested that the applied electric field during EPD is just responsible for the movement of particles towards the electrode, which then settle as a result of the pressure put forth by those incoming and in the outer layers. Böhmer showed that complex interplay of ionic, electrochemical, electroosmotic and hydrodynamic forces induces formation of densely packed sediment [44].

On the basis of our data collected so far, it can be deduced that the deposition on the membrane fits the Hamaker and Verwey [43] mechanism better. This is in agreement with Sarkar and Nicholson [34] and our three-layer-battery experiments [33]. The questions of the causes, and the processes of charge redistribution of adsorbed polyelectrolyte macromolecules to form the neutral deposited particulate composite, still remain unresolved and require further investigation.

Modulated high-resolution TGA tests were carried out with the purpose of determining the relative content of polymers in the deposited films. Figures 2a and 2b show the TGA runs of neat PAA, PEI and Celgard. Branched PEI undergoes a complete two-stage decomposition at 211.7 and 314.8°C. The first 10% weight loss of polyacrylic acid is measured at 178.7°C. This polyelectrolyte then decomposes stepwise at 258.7 and 385.2°C. Noteworthy that at 500°C the PAA loses 80% of its initial mass. The membrane is stable up to 356.7°C, and completely decomposes at 424°C. Lithium iron phosphate and lithium titanate are known as cathode and anode materials of high thermal stability At 500°C, their weight loss does not exceed 0.6 and 1.5%, respectively (Fig.2a). While there is an overlapping of the decomposition steps of polyelectrolytes, the dW/dT plots enable evaluation of their relative content in the composite coatings. Analysis of the thermograms of the deposits reveals that both polyelectrolytes, used as charging agents, are incorporated into the composite samples upon EPD. Cathodically deposited on membrane LTO films contain 5.1% of PEI and 13% PAA, while anodically deposited LFP films contain 4.8% PAA (Fig.2b). The change of substrate (Ni vs. Celgard membrane) does not significantly influence the composition of the films.

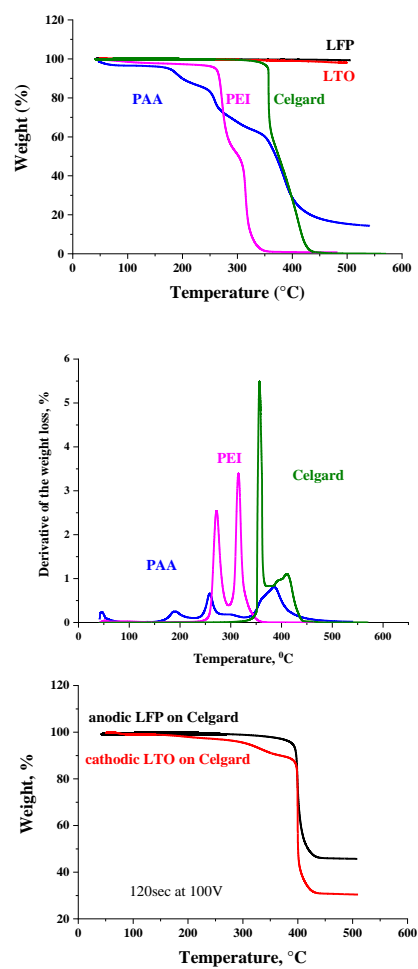

**Fig.2** TGA thermograms (a, c) and weight derivatives (b) of the active electrode materials, polyelectrolytes, Celgard, LTO and LFP deposits on the membrane

For the simultaneous deposition of LFP and LTO on the opposite sides of the membrane, the positive aluminum working electrode in our initial experiments was placed inside the heat-sealed bag prepared from Celgard 2325. The negative aluminum counter electrode was placed outside the Celgard bag in the homemade electrochemical-cell setup. The photo of this setup is shown in Figure 3a. The deposition bath and the holder for the electrodes were 3D-printed by the fused-filament-fabrication method. The holder, which has several gaps, enables control of the distance between the membrane and the electrodes and the suspension volumes in the device (Fig. 3b). The modified design of the homemade setup for simultaneous electrophoretic deposition on membrane, is shown in Figures 3c and 3d.

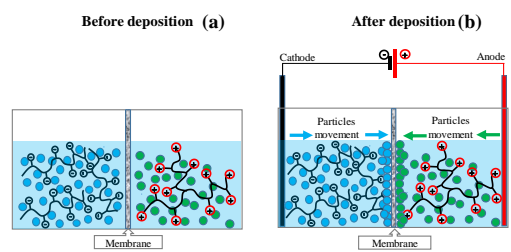

**Fig.4** Schematic presentation of the simultaneous EPD process before (a) and after application of electric field (b)

An optical photograph (Fig.5a) and cross-sectional ESEM image (Fig.5b) show the three-layer LFP/Celgard2335/LTO membrane electrode battery assembly prepared by EPD. It is worth mentioning, that no penetration of LFP and LTO particles via the membrane was detected at simultaneous EPD.

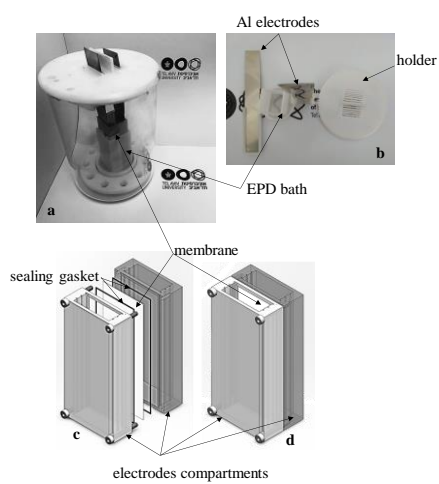

**Fig.3** Optical images (a, b) and schematics (c) of the home-made setups for the simultaneous EPD of electrodes on the opposite sides of the nanoporous membrane

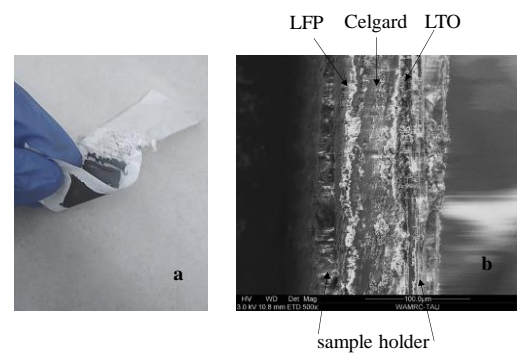

**Fig.5** Optical image (a) and cross-sectional ESEM view of the membrane electrode battery assembly prepared by simultaneous EPD

The acetone-based suspensions, comprising negatively charged LTO particles and positively charged LFP particles, were introduced to the different compartments of the cell, as schematically shown in Figure 4a. As described by Sarkar and Nicholson [35], in order to achieve deposition on a porous membrane, an ionic path must be maintained; in other words, there must be sufficient wetting of the membrane by the electrolyte solvent. In the case of suspensions based on acetone and ethanol, a 2325 Celgard membrane achieves sufficient wetting. Upon application of external DC voltage to the cell, the concurrent EPD of oppositely charged particles starts immediately. Positively charged LFP particles migrate towards the cathode, while negatively charged LTO particles simultaneously migrate towards the anode. Neither LFP (~200nm PSD) nor LTO (~150nm PSD) can penetrate the membrane and, as a result, they coat the opposite sides of the membrane. A pictorial illustration of this process is shown in Figure 4b. At low deposition times (up to 120sec) the deposition rate of concurrently on-membrane-deposited materials is close to that of separately deposited LFP and LTO composites.

The ESEM micrographs of the single-side and double-side deposits are compared in Figure 6. At low magnification, the morphology of all the samples appears homogeneous (see insets). High-resolution ESEM enables the observation of mostly separated $LiFePO_4$ and LTO particles (Figs. 6a and 6b) in the deposits. A network of carbon, filling the gaps between LFP and LTO particles, can be easily distinguished at high-magnification. While solitary aggregates are detected in the cathodically deposited LFP and LTO films, the separation of the particles in anodic coatings of both materials is much better. It seems likely that the simultaneous deposition of both materials on the opposite sides of Celgard does not influence the morphology of the films.

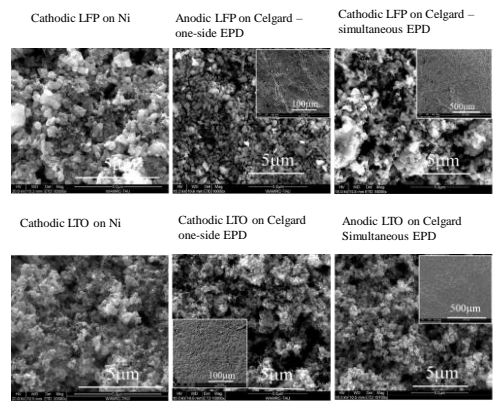

**Fig.6** Planar ESEM micrographs of the EPD films on Ni and Celgard

Time-of-Flight Secondary Ion Mass Spectrometry (TOF-SIMS) tests were performed both in the negative-ion and positive-ion modes with the goal of determining the spatial distribution of the active-material particles and polyelectrolytes in the samples (Fig. 7a–d). Brief Cs+ sputtering (cleaning of the surface) was used to obtain higher ionic yield and better resolution of the positive-ion images of the electrodes. Species of Li, Fe (from LiFePO$_4$), C, CHN and CHO (fragments from PEI and PAA) were detected in the positive-ion mass spectra acquired from the surface of the LFP electrode prepared by simultaneous EPD on one side of the membrane. The lowest measured intensity in the normalized individual-ion and total-ion images (10μm×10μm) corresponds to the darkest color and the highest intensity to the brightest one. Ion images (Fig. 7a) show that the pores between single particles and aggregates of the LFP cathode are filled with polymer, which also coats the LFP surface. Figure 7b shows good intermixing of both polyelectrolytes, PEI and PAA. The spatial distribution of lithium titanate and polyacrylic acid fragments in the anodically deposited LTO composite on the second side of the membrane resembles the ESM morphology of the sample (Fig. 7c) demonstrating porous morphology, as well. Cross-sectional TOFSIMS images Fig.7d support our optical and ESEM observations of the simultaneous deposition of LFP and LTO composite layers on the opposite sides of Celgard membrane and the absence of their intermixing. When imaging Fe in the LFP layer, there is some weak signal which is seen also on LTO side (green path on the top). It is due to the fact that there is a mass peak of C3H6N existing on Ti side, and it has a close mass to Fe, which is not real signal of the LFP. This effect is eliminated by narrowing the mass peaks for imaging. The right cross-sectional image in Fig.7d, which shows the mass-spectra signatures of PAA (red), overlapping of PAA and PEI (orange) and PP (green), provides evidence that Celgard membrane remains intact with the electrophoretically deposited electrodes

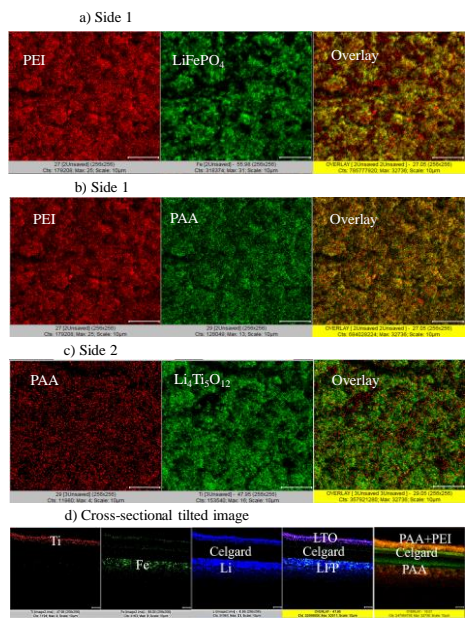

**Fig.7** Lateral, two-side and cross-sectional distribution of the components in the membrane electrode assembly

The results of electrochemical testing of half- and full-coin-cells (LFP/Li, LTO/Li and LFP/LTO) with the electrodes deposited electrophoretically on Celgard, are collected in Figure 8. The profiles of voltage vs. state-of-charge of lithium cells, containing LFP and LTO electrodes deposited on the membrane, are found to be similar to the cells with EPD electrodes deposited on nickel and to the voltage profiles of commercial batteries (Fig. 8a-c). The charge/discharge behavior of LFP/Li battery was thoroughly studied and described by Srinivasan and Newman [45]. The core/shell model suggests a biphasic structure of the LFP particle, a lithium-rich and a lithium-poor phase. The relative volume of each phase changes with the state of charge, and equilibrium between the phases is not reached within the time period of cycling. For lithiation, this causes a growing layer of fully lithiated LFP on the surface of the LFP particle followed by increased resistance to a further lithiation step. For delithiation, a lithium-depleted layer grows on the surface, increasing the extraction barrier and overpotential of lithium ion from within the particle [45, 46]. The overpotential growth is particularly pronounced at low state of charge, which is explained by the existence of narrow single-phase regions between a SOC of 0 and 0.02 and of 0.9525 and 1.0 [45]. At these SOC the insertion/extraction barrier of Li$^+$ to/from LFP is the highest one. The deposition of electrodes for the formation of MEBA was carried out at 100V and 30sec. The LFP/Celgard/LTO MEBA cells have been cycled at different C-rates with 125-140mAh/g capacity, which is close to that of the theoretical LFP value. Increasing the EPD electrodes thicknesses is expected to increase the capacity of MEBA cells. The plots recorded at cycle 1, 25,75 and 150 are identical with no evolution of an additional overpotential on prolonged cycling. While the charge/discharge overpotential of the cell with tri-layer LFP/Celgard/LTO MEBA and impregnated LiPF$_6$ EC:DEC electrolyte is 25mV, which is higher than that of the commercial cells, we are confident that optimization of the EPD process, namely the composition of the suspension and the deposition parameters, will enable unique, high-performance, flexible batteries by a one-step low-cost fabrication method. The simultaneous EPD can be used for the simple and low-cost fabrication of a variety of cathode (LMNO, NMC, NCA, $Li_2S$, etc.) and anode (graphite, silicon, alloys, etc.) materials on nanoporous polymer and ceramic ion-conducting membranes for energy-storage devices.

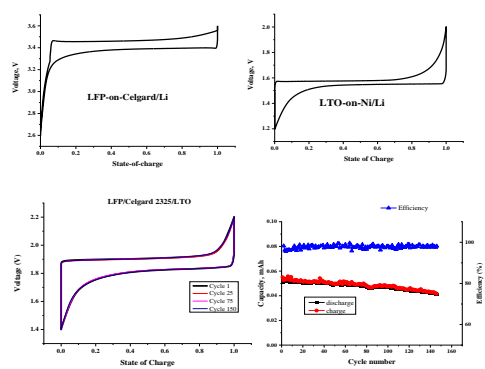

**Fig.8** Charge/discharge voltage profiles (a-c) of the cells with electrophoretically deposited on-membrane electrodes and cycle life (d) of the LFP/Celgard/LTO MEBA cell, comprising $LiPF_6$ EC:DEC electrolyte

## Conclusions

The novelty of the research presented in this manuscript is in the development of a prototype of flexible battery on nanoporous Celgard 2325 membrane. The LFP- and LTO-based electrodes were prepared by a single-step simultaneous electrophoretic deposition of oppositely charged particles. Both electrodes can be deposited anodically, or cathodically simply by the replacement of the charging agents. In the low-ionic-strength suspension the migration of both, positively charged particles and free ions (protons), followed by reduction of the latter is most likely to govern the cathodic EPD process. Passivation of the counter electrode (Ni or Al) by dissolved-in-acetone oxygen and migration of negatively charged particles control the anodic EPD. In the electrolytic bath with separate electrodes partitions, the nanoporous membrane sandwiched between the electrodes serves as the physical barrier to the two-way penetration of particles, the surface charge of which is induced by the adsorption of polyelectrolytes. Applied strong electric field facilitates coagulation of micelles in the close proximity to the membrane and precipitation of positively and negatively charged particles on its opposite sides. This is followed by the formation of tri-layer membrane electrode battery assembly. The lower EPD rate on the membrane than on metal electrodes can be explained by suppression of electric-field strength induced by polymer partition, and by partial blockage of the membrane pores.

PAA and PEI polyelectrolytes with carbon additives fill the cavities between the single particles and aggregates of the active materials in the electrophoretically deposited LFP and LTO-based battery electrodes. The LFP/Celgard/LTO MEBA cell has been reversibly cycled for more than 150 times.

## Experimental

*Preparation of the suspension and deposition process*

$LiFePO_4$ (LFP, Life Power P2, Clariant) and $Li_4Ti_5O_{12}$ (LTO, Life Power C-T2, Sud-Chemie Clariant) powders were used as active electrode materials. Super P C45 and C65 carbon (Timcal) - as conducting additives. Polyelectrolytes Poly(acrylic acid) (PAA) and Polyethyleneimine, branched PEI (25,000 MW, Sigma-Aldrich) - as the binders and charging agents. Commercial Celgard 2325 was used as a membrane.

For cathodic EPD, the PEI (0.1%) was used as a charging agent, and 2% PAA as a binder. For anodic EPD, the PAA (3% w/w) functions both as a charging agent and as a binder. For both anodic and cathodic EPD the active-material content (LFP or LTO) was 85%, carbon-additive content, 10%(w/w) and solid loading was 3-4mg/ml solvent. The suspension also included 0.25%(v/v) acetylacetone (Sigma-Aldrich).

At 30sec deposition time the standard deviation of the weight of the deposits was about 30%, at 60sec and at longer EPD process it decreased to 13.8% and 7.1%, respectively.

A Keithley SourceMeter model 2400 interfaced with LabTracer software and a PC was used to control the DC-EPD process and to monitor the current and voltage profiles. The constant voltage of 50 or 100V was applied to nickel or aluminum electrodes, in front of which the Celgard membrane was placed. The details of the setup are described in the "Results and Discussion" part of the manuscript.

The deposited three-layer thin-film LFP/Celgard/LTO samples were dried under vacuum at 50°C for 24 hours. The argon-filled MBraun glove box containing less than 0.1ppm water was used for handling of these materials.

*Cell assembly and characterization:*

The three-layer membrane-electrode battery assembly (MEBA) was soaked in commercial electrolyte (1M $LiPF_6$ in 1:1 EC:DEC:2%VC, Solvionics) and sealed in electrochemical coin cells (type 2032). The cells were cycled at room temperature in a Biologic VMP3 and BCS 805battery-test system.

A JSM-6300 scanning microscope (Jeol Co.) equipped with a Link elemental analyzer and a silicon detector, was used for the study of the surface morphology of the electrodes.

TOF SIMS tests were performed with the use of a TRIFT II (Physical Electronics Inc. USA) under the following operating conditions: primary ions (indium) , DC sputtering rate 0.035nm.min$^{-1}$ based on SiO$_2$ reference. Modulated high-resolution tests of the composite films were conducted with the use of high-sensitivity thermogravimetric analyzer Q5000 TGA-IR (TA Instruments), which operates from ambient temperature to 1000°C. The weight of the samples was 3-5mg.

## Conflicts of interest

There are no conflicts to declare.

## Acknowledgements

The authors would like to thank Prof. Y. Shapira (Emeritus in Department of Electrical Engineering-Physical Electronic, Tel Aviv University) and Dr. J. Penciner (School of Chemistry, Tel Aviv University) for the fruitful discussion of the manuscript. This work was partially supported by the Planning & Budgeting Committee of the Council of High Education and the Prime Minister Office of Israel, in the framework of the INREP project and by Israel Ministry of Science and Technology, project 3-12439.

## Notes and references